\documentclass[useAMS,usenatbib]{mn2e}
\usepackage{amssymb, latexsym, amsmath, graphicx, rotating, url, longtable}

\title[The AGN Outburst and Merger in HCG 62]{A Deep \emph{Chandra} Observation of the AGN Outburst and Merger in Hickson Compact Group 62}
\author[Rafferty et al.]{D.~A.~Rafferty,$^{1}$ L.~B\^{\i}rzan,$^{1}$ P.~E.~J.~Nulsen,$^{2}$ B.~R.~McNamara,$^{2, 3, 4}$
\newauthor
W.~N.~Brandt,$^{5}$ M. W. Wise$^{6}$ and H.~J.~A. R\"{o}ttgering$^{1}$ \\
$^{1}$Leiden Observatory, Leiden University, P.O.\ Box 9513, 2300 RA Leiden, The Netherlands\\
$^{2}$Harvard-Smithsonian Center for Astrophysics, 60 Garden St., Cambridge, MA 02138, U.S.A.\\
$^{3}$Department of Physics and Astronomy, University of Waterloo, Waterloo, N2L 2G1 Ontario, Canada\\
$^{4}$Perimeter Institute for Theoretical Physics, Waterloo, N2L 2YS Ontario, Canada\\
$^{5}$Department of Astronomy and Astrophysics, Pennsylvania State University, University Park, PA 16802, USA\\
$^{6}$ASTRON (Netherlands Institute for Radio Astronomy), P.O.\ Box 2, 7990 AA Dwingeloo, The Netherlands}
\begin{document}

\maketitle

\begin{abstract}
We report on an analysis of new \emph{Chandra} data of the galaxy group HCG 62, well known for possessing cavities in its intragroup medium (IGM) that were inflated by the radio lobes of its central active galactic nucleus (AGN). With the new data, a factor of three deeper than previous \emph{Chandra} data, we re-examine the energetics of the cavities and determine new constraints on their contents. We confirm that the ratio of radiative to mechanical power of the AGN outburst that created the cavities is less than 10$^{-4}$, among the lowest of any known cavity system, implying that the relativistic electrons in the lobes can supply only a tiny fraction of the pressure required to support the cavities. This finding implies additional pressure support in the lobes from heavy particles (e.g., protons) or thermal gas. Using spectral fits to emission in the cavities, we constrain any such volume-filling thermal gas to have a temperature $kT \gtrsim 4.3$~keV. For the first time, we detect X-ray emission from the central AGN, with a luminosity of $L_{2-10{\rm ~keV}} = (1.1 \pm 0.4)\times 10^{39}$~erg~s$^{-1}$ and properties typical of a low-luminosity AGN. Lastly, we report evidence for a recent merger from the surface brightness, temperature, and metallicity structure of the IGM.
\end{abstract}


\section{Introduction}
In the past decade, images from the \emph{Chandra} X-ray Observatory of the hot atmospheres of clusters and groups have been critical in furthering our understanding of AGN feedback. These images often show the radio lobes of the AGN displacing the hot gas, leaving cavities in the X-ray emission that are filled with radio emission \citep[e.g.,][]{mcna00, fabi00, blan01}. These images are direct evidence of the strong coupling between the radio lobes of the AGN and the hot, cooling gas. Estimates of the power required to produce the cavities has revealed that, in many such systems, the total jet power of the AGN is typically factors of 100 larger than the radiative power at monochromatic radio wavelengths \citep{birz04}. In massive cooling flows, the total jet power is on average similar to the total cooling luminosity of the hot gas \citep{raff06}, whereas in lower mass groups and ellipticals, it tends to be factors of several larger than the cooling luminosity, but with a lower duty cycle \citep[e.g.,][]{nuls09, dunn10, dong10}. Therefore, it is critical to study cavity systems in detail to understand fully the impact of the AGN on their group- and cluster-scale environments. In particular, the properties of AGN cavities at the group scale are less well known, due to the relatively few cases of clear AGN cavities in groups.

We report on the results of an analysis of deep \emph{Chandra} data on HCG 62, one of the best known cases of AGN feedback in a group. HCG 62 is a nearby ($z = 0.0137$), compact group of $\sim 60$ galaxies. Three large galaxies (NGC 4761, NGC 4776, and NGC 4778) dominate the core, with NGC 4778, an S0 galaxy, being the brightest cluster galaxy (BCG). The BCG hosts a low-luminosity AGN with weak line emission \citep{cozi98, shim00} and weak radio lobes \citep{gitt10, giac11}. Mid-infrared \emph{Sptizer} observations by \citet{gall08} show no evidence for excess nuclear mid-infrared emission, and no X-ray emission from the AGN was detected by \citet{mori06}, who found an upper limit for this emission of $L_{\rm 0.5-4~keV} \lesssim 10^{39}$~erg~s$^{-1}$. Optical observations of the BCG show no evidence of recent star formation \citep{raff08}. The group is X-ray luminous, with a 0.5--8 keV luminosity  of $\approx 10^{43}$~ergs~s$^{-1}$. The temperature and density profiles of the IGM \citep[e.g.,][]{mori06} show the classical cooling-flow behavior in which the temperature decreases inwardly to a cool, high-density core. The cooling time in the core is $\approx 6\times 10^{7}$~yr \citep{raff06}.

HCG 62 was first observed by \emph{Chandra} in 2000 for $\approx 50$~ks. This observation revealed cavities in the IGM that were subsequently shown to be filled with radio emission from the lobes of the central AGN \citep{vrti02, birz04, mori06, gitt10, giac11}. Interpreting these cavities as buoyant bubbles, \citet{birz04} calculated that the cavities represent a total mechanical power of $\sim 3 \times 10^{42}$~erg~s$^{-1}$, more than enough to offset cooling losses of the IGM in the core. HCG 62 has also been observed by XMM-Newton \citep{mori06, gitt10}, ASCA \citep{naka07}, and Suzaku \citep{toko08}, providing a wealth of X-ray data at a variety of scales, resolutions, and energies. Therefore, HCG 62 is an ideal candidate for detailed studies of various properties of the hot atmosphere in groups.

We report here on new, deep \emph{Chandra} observations of HCG 62 that reveal further cavities and evidence of a recent merger. We adopt $H_0 = 70$~km~s$^{-1}$~Mpc$^{-1}$, $\Omega_{\Lambda} = 0.7$, and $\Omega_{\rm{M}} = 0.3$, implying a luminosity distance to HCG 62 of 59.3~Mpc and a angular scale of 0.28~kpc~arcsec$^{-1}$.

\section{Data Reduction and Analysis}\label{S:data_reduction}
HCG 62 was observed by the \emph{Chandra} X-ray Observatory on 2000/01/25 for 48.5~ks (ObsID 921, FAINT mode), on 2009/03/02 for 67.1~ks (ObsID 10462, VFAINT mode), and on 2009/03/03 for 51.4~ks (ObsID 10874, VFAINT mode), all with the \mbox{ACIS-S} detector. The data were reprocessed with \textsc{CIAO} 4.2\footnote{See \url{cxc.harvard.edu/ciao/index.html}.} using \textsc{CALDB} 4.2.1\footnote{See \url{cxc.harvard.edu/caldb/index.html}.} and were corrected for known time-dependent gain and charge transfer inefficiency problems.  For each observation, blank-sky background files were used for background subtraction.\footnote{See \url{http://asc.harvard.edu/contrib/maxim/acisbg/}.} The events files were filtered for flares using the \textsc{CIAO} script \emph{lc\_clean} to match the filtering used during the construction of the background files. A total of 7.5 ks was removed during filtering, resulting in a combined exposure time of 159.5 ks. The background files were normalized to the count rate of the source image in the $10-12$ keV band (after filtering). The derived normalization ratios (in the sense of background/source) were 1.01 for ObsID 921 and 0.92 for both ObsID 10462 and ObsID 10874. Lastly, point sources detected using the \textsc{CIAO} tool \emph{wavdetect} were removed.

\subsection{Imaging Analysis}\label{S:imaging_analysis}
Images used in the analysis that follows were made with the filtered events data. To correct for exposure, which is a function of both position and energy, images were made in 12 narrow energy bands from 0.5~keV to 7~keV. The bands were chosen to enclose energy ranges for which the ACIS-S detector has approximately constant response. Exposure maps were weighted for each narrow-band image using a monochromatic energy equal to the central energy of each band. The narrow-band images and exposure maps from the three observations were reprojected to the same grid and summed together using the CIAO tool \emph{reproject\_image}. Regions where point sources were removed from the events data were filled with background emission using the \textsc{CIAO} tool \emph{dmfilth}. The narrow-band images were then divided by their associated exposure map and summed together to produce the final, broad-band (0.5--3~keV), exposure-corrected image used in subsequent imaging analysis.

\begin{figure*}
\includegraphics[width=168mm]{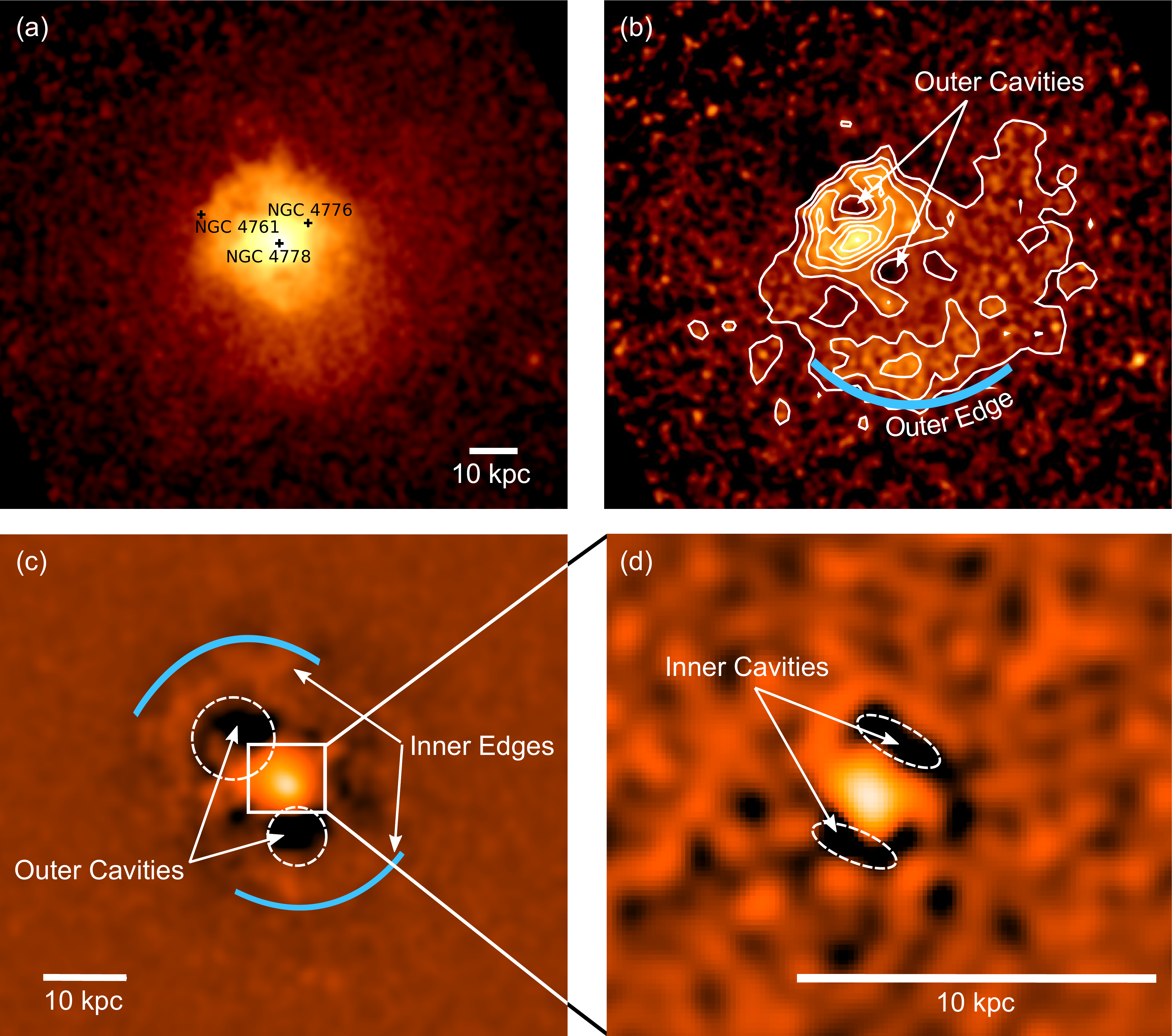}
\caption{\emph{(a)} Exposure-corrected, 0.5--3~keV image of HCG 62, smoothed with a gaussian of $\sigma=5$\arcsec. \emph{(b)} Residual image (in the sense of [model - image] / model) with contours, made using the exposure-corrected 0.5--3~keV image shown in panel a. The model was made using a Multi-Gaussian Expansion \citep{capp06}. The linear scale is identical to that of panel a. \emph{(c)} Unsharpmask image of the core made by subtracting an image smoothed at 32\arcsec\ from an image smoothed at 6\arcsec. \emph{(d)} Unsharpmask image of the inner core made by subtracting an image smoothed at 5\arcsec\ from an image smoothed at 4\arcsec.}
\label{F:fullband_image}
\end{figure*}

The exposure-corrected 0.5--3~keV image is shown in Figure~\ref{F:fullband_image}a. A number of features noted by other authors \citep[e.g.,][]{mori06,gitt10} in analyses of the original data are apparent, including the large cavities, a sharp inner surface brightness edge just beyond the cavities, and an outer surface brightness edge some 30~kpc ($\sim 2$~arcmin) to the south of the core.

\begin{figure}
\includegraphics[width=84mm]{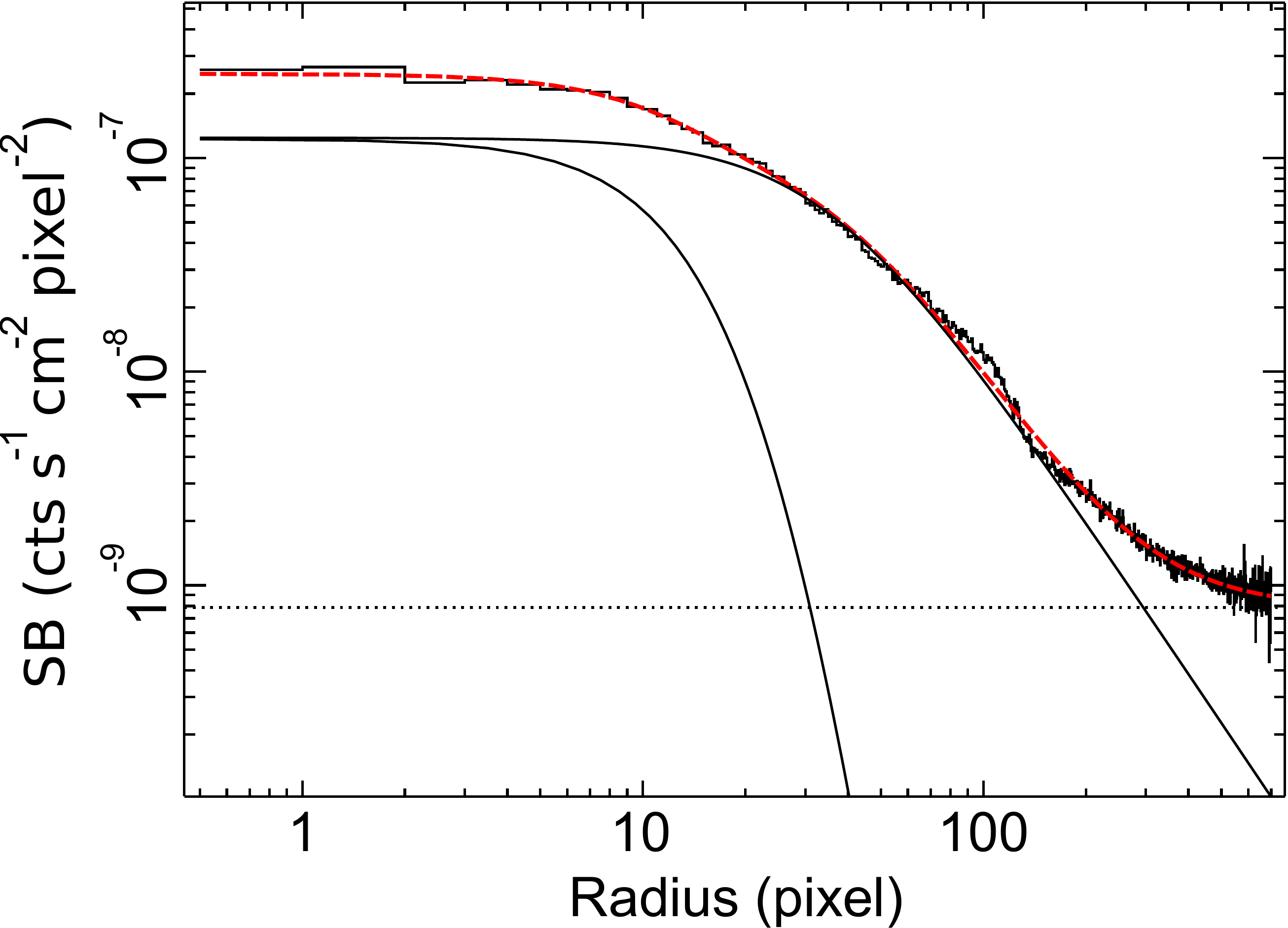}
\caption{The azimuthally averaged surface brightness profile (0.5--3~keV), with the best-fitting double-$\beta$ model (solid lines) plus constant background model (dotted line) overplotted. The sum of the $\beta$ model and background components is shown by the dashed line. The sharp inner edge is visible at a radius of $\sim 120$ pixels ($\approx 1$~arcmin).  \label{F:sb_profile_beta}}
\end{figure}

We derived an azimuthally averaged surface brightness profile from the exposure-corrected 0.5--3~keV image by extracting the mean surface brightness in 1-pixel-wide circular annuli. No masking of the features mentioned above was done. The resulting profile is shown in Figure~\ref{F:sb_profile_beta}. To parametrize the profile, we fit a double-$\beta$ model \citep{sara77} with a constant background to the profile. The best-fitting parameters are given in Table~\ref{T:beta_model}. Although this model fits well over most of the profile, a clear departure from the data is apparent at a radius of $\sim 120$ pixels ($\sim 60$\arcsec), which corresponds to the location of the inner surface brightness edge. We find no evidence for a point-like source, indicative of X-ray emission from an AGN, at the centre of the profile, confirming the findings of \citet{mori06}.

We also fit a 2-dimensional circular double-$\beta$ model to the core region only ($r<15$~kpc). We allowed the positions, normalizations, and $\beta$ parameters of the $\beta$-model components to vary freely. Fitting was performed in Sherpa version 4.2\footnote{See \url{http://cxc.harvard.edu/sherpa/index.html}.} to the exposure-corrected 0.5--3~keV image, with the cavity regions masked. The best-fitting parameters are given in Table~\ref{T:beta_model}. These parameters differ somewhat from those determined from the large-scale azimuthally averaged surface brightness profile, but provide a better fit to the core emission. This 2-dimensional model is later used in the determination of the cavity sizes (see \S\ref{S:energetics}).

\begin{table*}
\begin{minipage}{126mm}
\caption{Parameters of the $\beta$-model Fits.}
\label{T:beta_model}
\begin{tabular}{@{}lccccc}
\hline
  & \multicolumn{2}{c}{Centre Position (J2000)} &  &  & Core Radius \\
Comp. & RA & Dec & Norm$^a$ & $\beta$ & (pixels) \\
\hline
\multicolumn{6}{c}{Azimuthally Symmetric Model} \\
\hline
$\beta_1$ & 12:53:05.7 & -09:12:14.1 & $(1.24 \pm 0.01) \times 10^{-7}$  & $0.563  \pm 0.001$  &  $35.3 \pm 0.1$ \\
$\beta_2$ & 12:53:05.7 & -09:12:14.1 & $(1.23 \pm 0.04) \times 10^{-7}$  & $2.23   \pm 0.07$ & $27.5   \pm 0.5$ \\
BG        & \ldots    & \ldots     & $(7.87 \pm 0.05) \times 10^{-10}$ & \ldots    & \ldots \\
\hline
\multicolumn{6}{c}{2-D Model} \\
\hline
$\beta_1$ & 12:53:05.7 & -09:12:14.6 & $(2.74 \pm 0.13) \times 10^{-7}$ & $0.33 \pm 0.05$ & $6.2  \pm 0.7$ \\
$\beta_2$ & 12:53:06.0 & -09:12:09.7 & $(9.0  \pm 0.65) \times 10^{-8}$ & $0.60 \pm 0.04$ & $44.2 \pm 2.5$ \\
\hline
\end{tabular}
\medskip
$^a$Normalizations are for the 0.5--3~keV band and have units of counts s$^{-1}$ cm$^{-2}$ pixel$^{-2}$.
\end{minipage}
\end{table*}

To minimize artefacts due to asymmetries in the large-scale background group emission that are not accounted for in the $\beta$-model fits, we also modeled the exposure-corrected 0.5--3~keV image using the Multi-Gaussian Expansion method of \citet{capp06}. Although this method was developed for use with optical images of galaxies, we have found it to work well with X-ray images that contain a large number of counts. We restricted the axial ratio of Gaussians to be between 0.2 and 1 to exclude highly elliptical Gaussians that are not appropriate for the smooth, large-scale background group emission. The residual image for this model highlights asymmetries in the distribution of the hot gas and is shown in Figure~\ref{F:fullband_image}b.

Lastly, to highlight the cavities and surface-brightness edges, we made unsharp-mask images at two scales: an image at large scales was made by subtracting an image smoothed at 32\arcsec\ from an image smoothed at 6\arcsec, and an image at small scales was made by subtracting an image smoothed at 5\arcsec\ from an image smoothed at 4\arcsec. These images are shown in Figures~\ref{F:fullband_image}c and \ref{F:fullband_image}d.

There are a number of features in these images that were not readily apparent in the original, shallower \emph{Chandra} data. First, on large scales, one can see a spiral-like region of excess emission that extends to the south of the core, which has an outer boundary approximately coincident with the surface brightness edge in that region. The contours shown in Figure~\ref{F:fullband_image}b highlight this excess. On smaller scales, in addition to the large cavities found previously, two small cavities are now apparent on either side of the core, oriented roughly north-south. These features will be discussed further in \S\ref{S:cavities} and \S\ref{S:merger}.


\subsection{Spectral Analysis}\label{S:spectral_analysis}
For the spectral analysis, spectra were extracted from each observation separately using the \textsc{CIAO} script \emph{specextract}. For each spectrum, weighted responses were made, and a background spectrum was extracted in the same region of the CCD from the associated blank-sky background file. Joint spectral fitting, using all three observations, was then performed when deriving spectral properties. For the spectral fitting, \textsc{XSPEC} \citep{arna96} version 12.5.1 and \textsc{Sherpa} version 4.2 were used.

\subsubsection{Azimuthally Averaged Properties}\label{S:deprojection}
X-ray spectra were extracted in circular annuli centred on the centroid of the cluster emission. The annuli were constructed from ObsID 921 such that each annulus contained $\approx 4500$ counts, with the exception of the central annulus, which contained $\approx 1000$ counts (to allow derivation of the various properties as close to the core as possible). Spectra were then extracted from ObsIDs 10462 and 10874 from the same regions and generally contained a similar number of counts. Gas temperatures and densities were found by deprojecting these spectra using the Direct Spectral Deprojection method of \citet{sand07}.\footnote{See \url{http://www-xray.ast.cam.ac.uk/papers/dsdeproj/}.} The deprojection was done for each observation separately. The three deprojected spectra in each annulus (one for each of the observations) were then fit jointly in \textsc{XSPEC} with a single-temperature plasma model (MEKAL) absorbed by foreground absorption model (WABS), between the energies of 0.5 keV and 7.0 keV. In this fitting, the redshift was fixed to $z=0.0137$, and the foreground hydrogen column density was fixed to $N_{\rm H} = 3.03 \times 10^{20}$~cm$^{-2}$, the weighted-average Galactic value from \citet{dick90}.\footnote{A fit was also done with the foreground column density allowed to vary, but the differences in the best-fitting parameters were negligible. We also note that the value of $N_{\rm H}$ determined by \citet{kalb05} is $\approx 10$ per cent higher ($N_{\rm H} = 3.32 \times 10^{20}$~cm$^{-2}$). Fit results did not change significantly with $N_{\rm H}$ fixed to this value.}

The density was then calculated from the normalization of the MEKAL component, assuming $n_{\rm{e}}=1.2n_{\rm{H}}$ (for a fully ionized gas with hydrogen and helium mass fractions of $X=0.7$ and $Y=0.28$).  The pressure in each annulus was calculated as $P=nkT,$ where we have assumed an ideal gas and $n \approx 2n_{\rm{e}}$. The entropy is defined as $S=kTn_{\rm e}^{-2/3}$. The cooling time was derived from the temperature, metallicity, and density using the cooling curves of \citet{smit01}, calculated using \textsc{APEC}.

\begin{figure*}
\includegraphics[width=168mm]{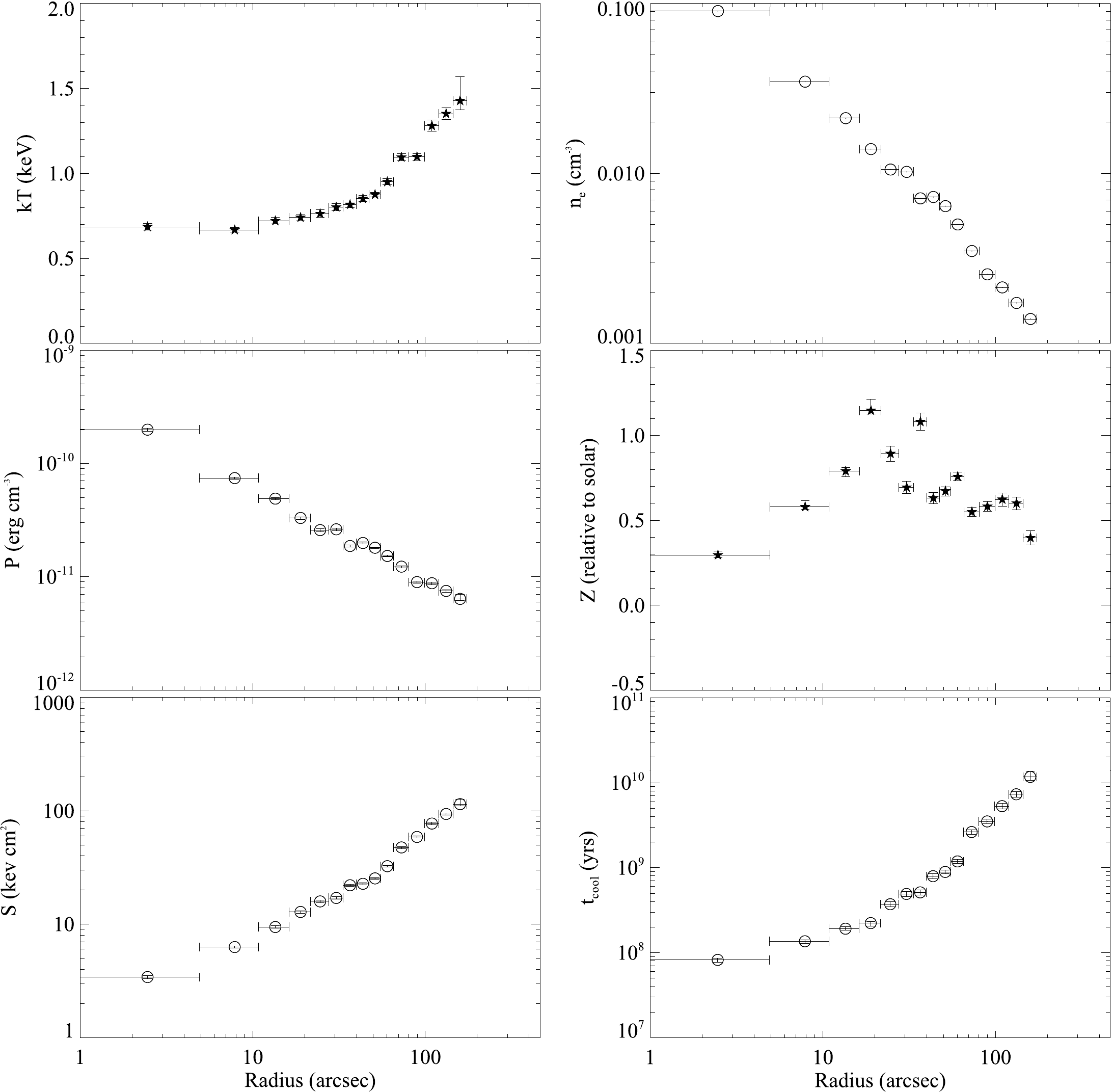}
\caption{Radial profiles of temperature, density, pressure, metallicity, entropy, and cooling time (note that 10\arcsec\ equals 2.8~kpc).}
\label{F:profiles}
\end{figure*}

The resulting radial profiles of temperature, density, pressure, metallicity, and cooling time are shown in Figure~\ref{F:profiles}. A number of features of these profiles are noteworthy. First, there appear to be two clear discontinuities in the temperature profile at $r\approx 60$\arcsec\ and at $r\approx 90$\arcsec. The relation of these discontinuities to the surface brightness edge seen at a similar radius in Figure~\ref{F:sb_profile_beta} is discussed further in \S\ref{S:inner_sb_edge}.

The second notable feature in the radial plots is an off-centre peak (or, equivalently, a central dip) in the metallicity profile (this dip is also visible in the temperature map discussed in \S\ref{S:maps}). The metallicity has a value of $\approx 0.3$ solar in the innermost bin, rising to approximately solar metallically at a radius of 20\arcsec--30\arcsec, then dropping to one-half solar at the group outskirts. Similar features have been noted in a number of groups and clusters \citep[e.g., the Centaurus Cluster,][]{sanders02}. There has been some debate in the literature as to whether such features are real or not, and a number of possibilities have been proposed to explain them, such as resonance scattering \citep[e.g.,][]{sand06a} and modeling effects due to the presence of multi-phase gas \citep[e.g.,][]{buot98,buot00,mole01}.

Resonance scattering has been found to be generally insufficient to account for the observed metallicity deficits \citep{sand06a}, and we do not investigate this possibility further. Instead, we examine the possibility that multi-phase gas in the core of HCG 62 has resulted in metallicities that are biased low. To this end, we followed the analysis of \citet{kirk09} and added in the inner regions a multi-phase cooling-flow component (MKCFLOW) with a low temperature fixed to 0.1~keV. The metallicity and high temperature parameters of the cooling-flow component were allowed to vary freely. Although an F-test does not indicate that the cooling-flow component improves the fit significantly, we do find that the metallically of the single-temperature component increases in the inner three regions with the addition of the cooling-flow component (see Table~\ref{T:metallicities} and Figure~\ref{F:profiles}), largely erasing the central metallicity dip. This result is true for fits to both the projected and deprojected spectra, although the deprojected spectra show smaller metallicities in general (note, too, that the models generally fit much better to the deprojected spectra than to the projected ones). Therefore, it is possible that the central metallicity dip is due to our modeling of a multi-phase gas in the core with a simple single-temperature model.
\begin{table}
\caption{Parameters of the Spectral Fits in the Core.}
\label{T:metallicities}
\begin{tabular}{@{}lcccc}
\hline
  & \multicolumn{2}{c}{Metallicity} & \multicolumn{2}{c}{$\chi^2_{\nu}$ (dof)} \\
Annulus$^a$ & 1T & 1T+CF & 1T & 1T+CF \\
\hline
\multicolumn{5}{c}{Projected Spectra}\\
\hline
    1 & $0.49_{-0.10}^{+0.07}$ & $1.20_{-0.26}^{+1.84}$ & 1.20 (672) & 1.09 (669) \\
    2 & $0.60_{-0.08}^{+0.12}$ & $1.63_{-0.54}^{+0.84}$ & 3.38 (670) & 3.31 (667) \\
    3 & $0.95_{-0.12}^{+0.20}$ & $2.42_{-0.65}^{+2.24}$ & 5.17 (670) & 5.04 (667) \\
    \hline
\multicolumn{5}{c}{Deprojected Spectra}\\
\hline
    1 & $0.30_{-0.01}^{+0.02}$ & $0.93_{-0.21}^{+0.36}$ & 1.45 (121) & 1.29 (118) \\
    2 & $0.58_{-0.03}^{+0.03}$ & $1.11_{-0.38}^{+0.71}$ & 1.26 (121) & 1.25 (118) \\
    3 & $0.79_{-0.02}^{+0.03}$ & $1.68_{-0.43}^{+1.44}$ & 1.20 (121) & 1.13 (118) \\
\hline
\end{tabular}
\medskip
$^a$See Figure~\ref{F:profiles} for inner and outer radii of the annuli.
\end{table}

\subsubsection{Temperature and Abundance Maps}\label{S:maps}
With the large number of counts available with the new data, it is useful to derive 2-dimensional maps of temperature and metallicity. To this end, bins were constructed for spectral extraction from the raw 0.5--7~keV counts image shown in Figure~\ref{F:fullband_image} using two binning algorithms: the Contbin algorithm of \citet{sand06b}, which constructs bins that roughly follow contours in the counts image, and the \textsc{WVT} algorithm of \citet{dieh07}, which constructs bins using Voronoi tessellation. The bins were constructed to contain at least 2000 counts each (totaled over all three observations). Spectra and responses were extracted from each observation separately and were then fit jointly with an absorbed single-temperature model in \textsc{Sherpa}. As before, the absorbing column density was fixed to the foreground hydrogen column density of $N_{\rm H} = 3.03 \times 10^{20}$~cm$^{-2}$, and the redshift was fixed to $z=0.0137$.

\begin{figure*}
\begin{tabular}{@{}cc}
\includegraphics[width=84mm]{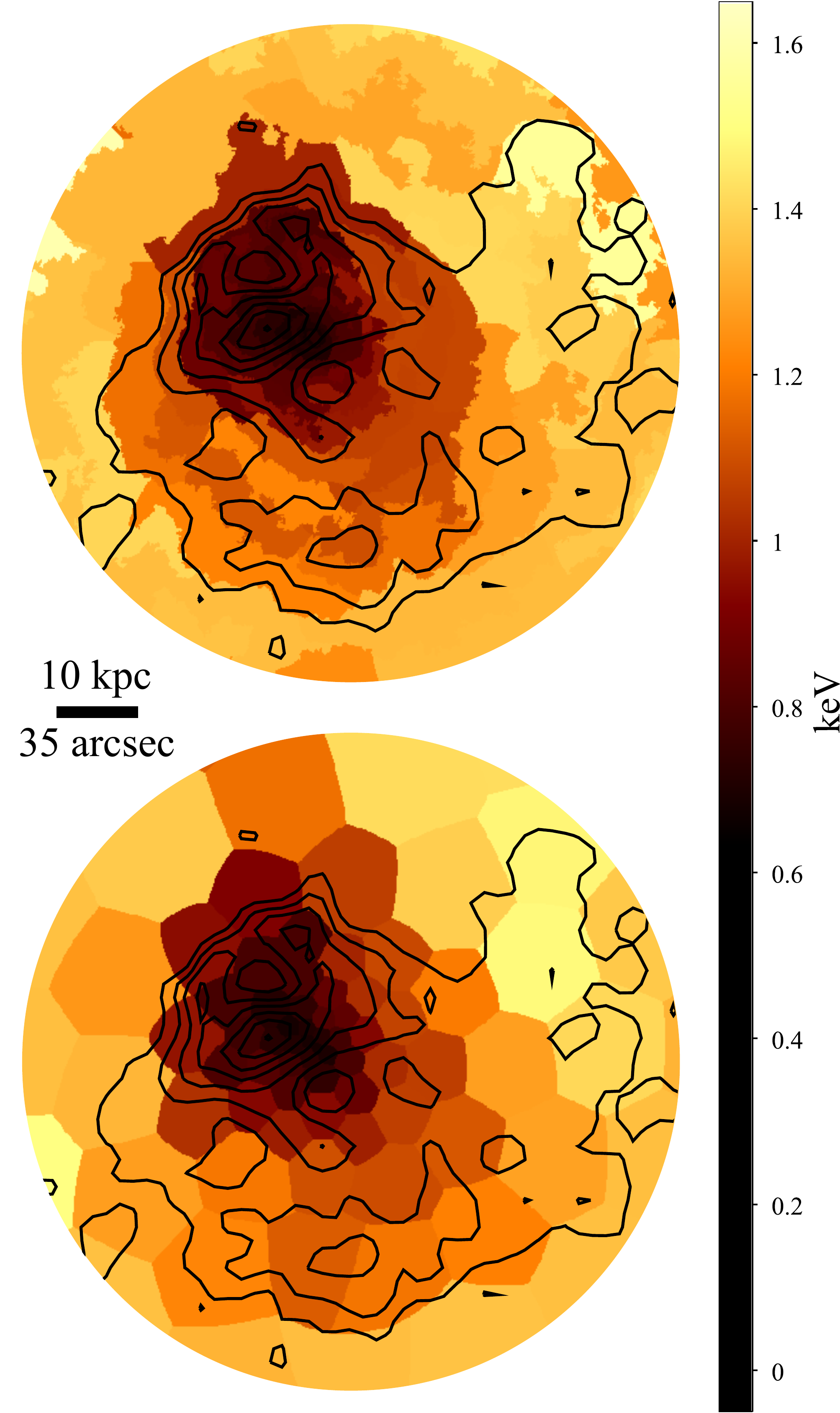} &
\includegraphics[width=84mm]{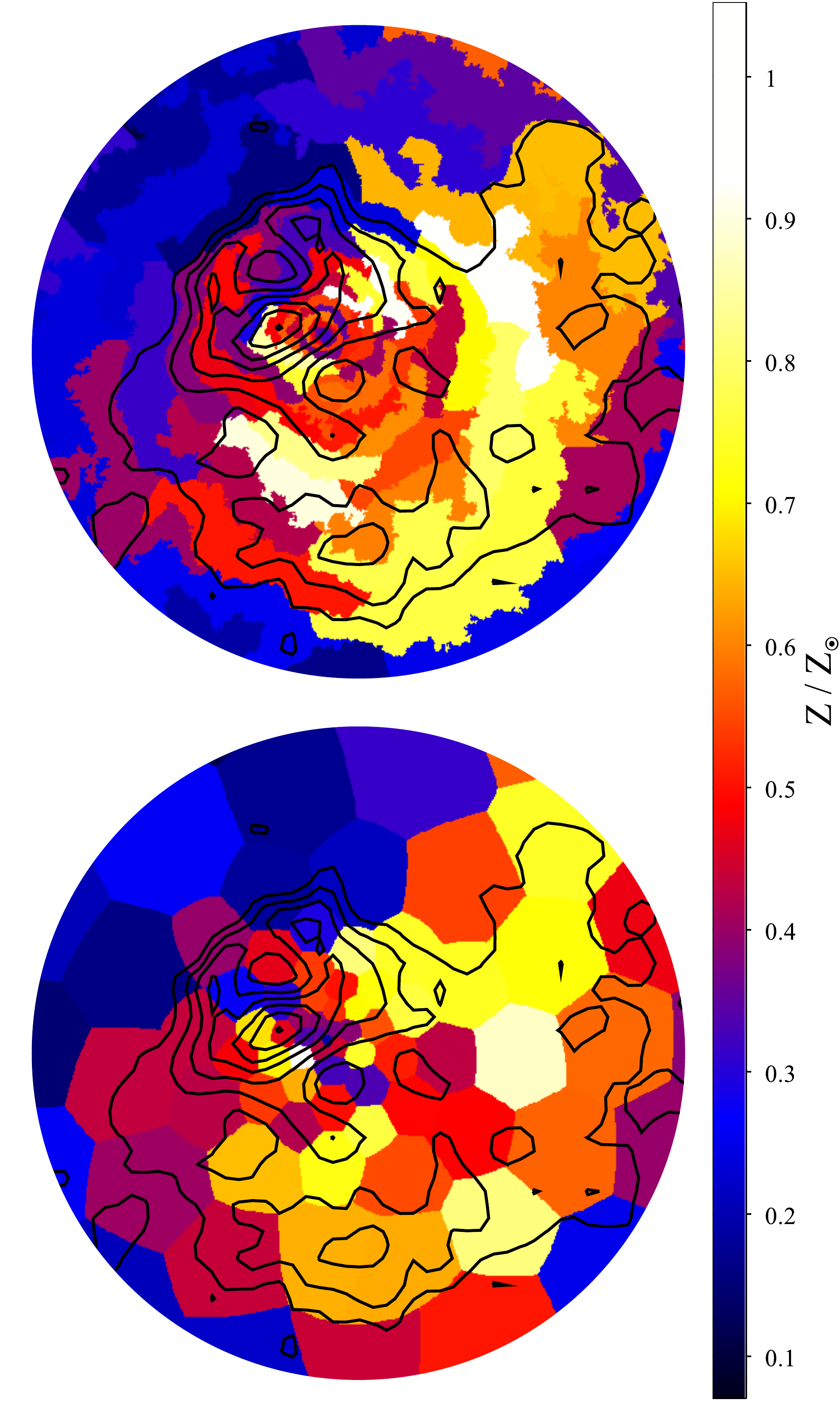} \\
\end{tabular}
\caption{\emph{Left:} Temperature maps made using contour binning (\emph{top}) and using WVT binning (\emph{bottom}) with contours from the residual image (Figure~\ref{F:fullband_image}) overlaid. Errors on the values in each bin are typically 2\%--4\%. Note the presence of cooler gas to the inside of the surface-brightness edge (indicated by the southern-most contour)  and warmer gas to the outside, a typical signature of a cold front. \emph{Right:} Metallicity maps (relative to solar) with contours from the residual image overlaid. Errors are typically 20\%--30\%.}
\label{F:maps}
\end{figure*}

The resulting maps of temperature and metallicity for each binning scheme are shown in Figure ~\ref{F:maps}. Beyond the general trends seen in the radial temperature and metallicity profiles of higher temperatures and lower metallicities with increasing distance from the core, there is a clear asymmetry in both maps, with low-temperature, high-metallicity gas extending further to the south of the core than to the north. In the metallicity map, the asymmetry appears as a high-metallicity arc extending to the south and west of the core. This arc is discussed further in Section~\ref{S:arc}.

\subsection{Nuclear Point Source Emission}
NGC 4778, the BCG of HCG 62, is thought to host a low-luminosity AGN (LLAGN) at its core \citep{cozi98, shim00}. \citet{mori06} found no evidence for point-like X-ray emission at the location of the optical core of NGC 4778 ($\alpha_{\rm J2000} = 193.2738^{\circ}$, $\delta_{\rm J2000} = -9.2040^{\circ}$) reported by \citet{zabl00}. The new data confirm this finding. However, \citet{raff08} report a somewhat different location for the optical core ($\alpha_{\rm J2000} = 193.2738^{\circ}$, $\delta_{\rm J2000} = -9.2040^{\circ}$), differing by $\approx 2$~arcsec from that reported by \citet{mori06}. The reason for this difference is unclear. However, the \citet{raff08} location coincides precisely with possible point-like emission in the 2--7~keV band. To understand whether this emission could include a contribution from an AGN, we performed spectral fitting as follows.

A spectrum was extracted from each observation in a circular region of radius 0.83\arcsec\ centred on the location of the optical core of NGC 4778 as reported by \citet{raff08}. The region was sized to enclose 90 per cent of the energy of an unresolved source at 1~keV at the position on the CCD of NGC 4778. As before, we extracted background spectra from the appropriate blank-sky background files in the same 0.83\arcsec\ region.

The spectra from all three observations were fit simultaneously between 0.5--7~keV in \textsc{XSPEC} with a model that included a thermal component (to account for IGM emission) and an absorbed power-law component (to account for AGN emission). As before, a WABS component was also used to account for Galactic absorption with the column density fixed to $3.03\times 10^{20}$~cm$^{-2}$. A ZWABS component was used to account for absorption intrinsic to the AGN. The model was then defined as WABS$\times$(MEKAL+ZWABS$\times$POW). Due to the low number of total counts (237, of which 4 are background; in the 2--7~keV band, there are 22 source counts and 1 background count), the Cash statistic was used on spectra binned to a minimum of 3 counts per bin (poor results were obtained when unbinned spectra were used). The power-law index was fixed to 1.8, typical of low-luminosity AGN \citep[e.g.,][]{ho01,tera03}. The metallicity of the MEKAL component was fixed to the value derived for the central bin from deprojection (0.3~$Z_\odot$). The temperature and normalization of the MEKAL component were allowed to vary, as was the column density of the ZWABS component. The redshifts of the MEKAL and ZWABS components were fixed to $z=0.0137$.

Our fits confirm the presence of an X-ray AGN at the core of HCG 62. From the best-fitting normalization of the power-law component, we find an implied intrinsic X-ray luminosity for the AGN of $L_{0.5-7{\rm ~keV}} = (1.5_{-1.0}^{+2.8})\times 10^{39}$~erg~s$^{-1}$ (3-$\sigma$ errors are quoted). The intrinsic column density is only poorly constrained, with the best-fitting value being $N_{\rm H} = 0.6_{-0.5}^{+1.1} \times 10^{22}$~cm$^{-2}$ (1-$\sigma$ errors) and $N_{\rm H} < 6.0 \times 10^{22}$~cm$^{-2}$ (3-$\sigma$ upper limit).
Therefore, the spectrum shows evidence for moderate absorption, as seen in a number of LLAGN \citep[e.g.][]{tera03}.

Using the H$\alpha$ flux implied by the measurements of \citet{cozi98} and \citet{shim00} for the nucleus of NGC 4778 ($F_{{\rm H}\alpha} = 7.3 \times 10^{-16}$~erg~cm$^{-2}$~s$^{-1}$), we can compare the ratio of the X-ray luminosity to the H$\alpha$ luminosity to that of typical LLAGN. Using the 2--10~keV luminosity implied by our fit of $L_{2-10{\rm ~keV}} = (1.1 \pm 0.4)\times 10^{39}$~erg~s$^{-1}$, we find a ratio of $L_{2-10 {\rm ~keV}} / L_{{\rm H}\alpha} \approx 2.7 \pm 1.0$, consistent with typical ratios in LLAGN of $\sim 1$--2 \citep[e.g.,][]{tera00, ho01}.

Lastly, we find evidence for variability of the 2--7 keV nuclear X-ray emission over the period spanned by our observations ($\sim 9$~yr). We find background-subtracted count rates in the 2--7~keV band of $(4.4\pm 3.1) \times 10^{-5}$~cnts~s$^{-1}$ for ObsID 921 \citep[taken in early 2000; errors are calculated following][]{gehr86} and $(15\pm 5) \times 10^{-5}$~cnts~s$^{-1}$ and $(21\pm 7) \times 10^{-5}$~cnts~s$^{-1}$ for ObsIDs 10462 and 10874, respectively (both taken in early 2009). Therefore, it appears the AGN has brightened by a factor of $\approx 3$ during this 9-year period.

\section{The AGN Outburst}\label{S:cavities}
As previous studies of HCG 62 have reported, the AGN has interacted strongly with the IGM in the core of the group. The large outer cavities visible in Figure~\ref{F:fullband_image}c are the most prominent feature of this interaction, but there is also the inner surface brightness edge noted in \S\ref{S:imaging_analysis} that, given its proximity to the cavities, is likely associated with them.

In this section, we examine the structure, energetics, and content of the cavity system. We first examine the surface brightness edge that encompasses the outer cavities to determine whether this feature could be due to a weak shock (which would contribute to the total energetics of the outburst) or might instead be due to rims of cool material surrounding the cavities such as those as seen in A2052 \citep{blan03}. We then estimate the total outburst energetics and place limits on the contents of the cavities.

\subsection{The Surface Brightness Edge}\label{S:inner_sb_edge}
\begin{figure}
\includegraphics[width=84mm]{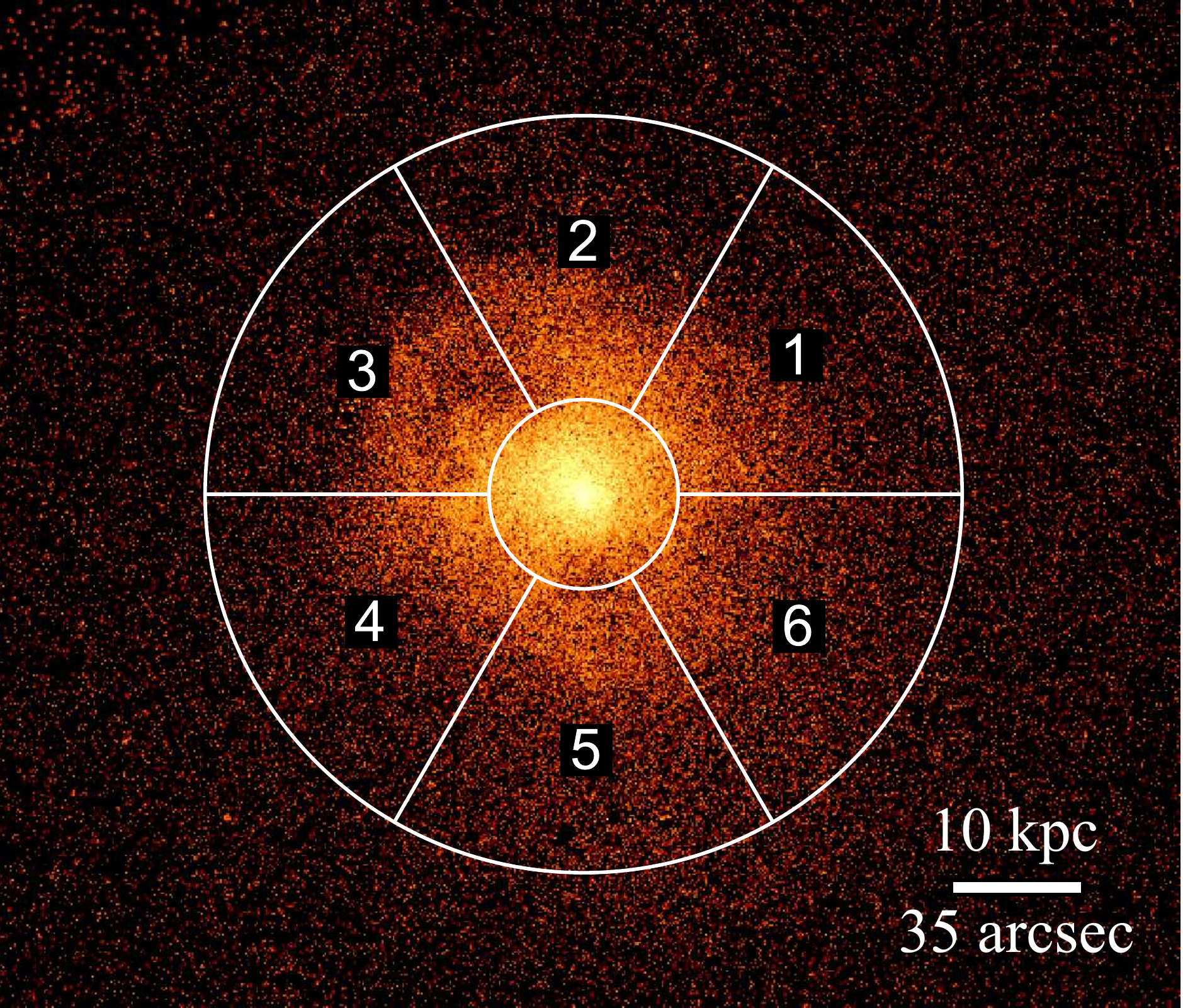}
\caption{The exposure-corrected 0.5--3~keV counts image with the six sectors used for fitting of the shock models to the the inner surface brightness edge. The inner and outer boundaries of the sectors correspond to 50 and 200 pixels (25\arcsec\ and 98\arcsec), respectively, the approximate range over which fitting of the inner edge was done (fitting of the outer edge was done in sector 5 between 120 and 500 pixels).}
\label{F:shock_regions}
\end{figure}

\begin{figure}
\includegraphics[width=84mm]{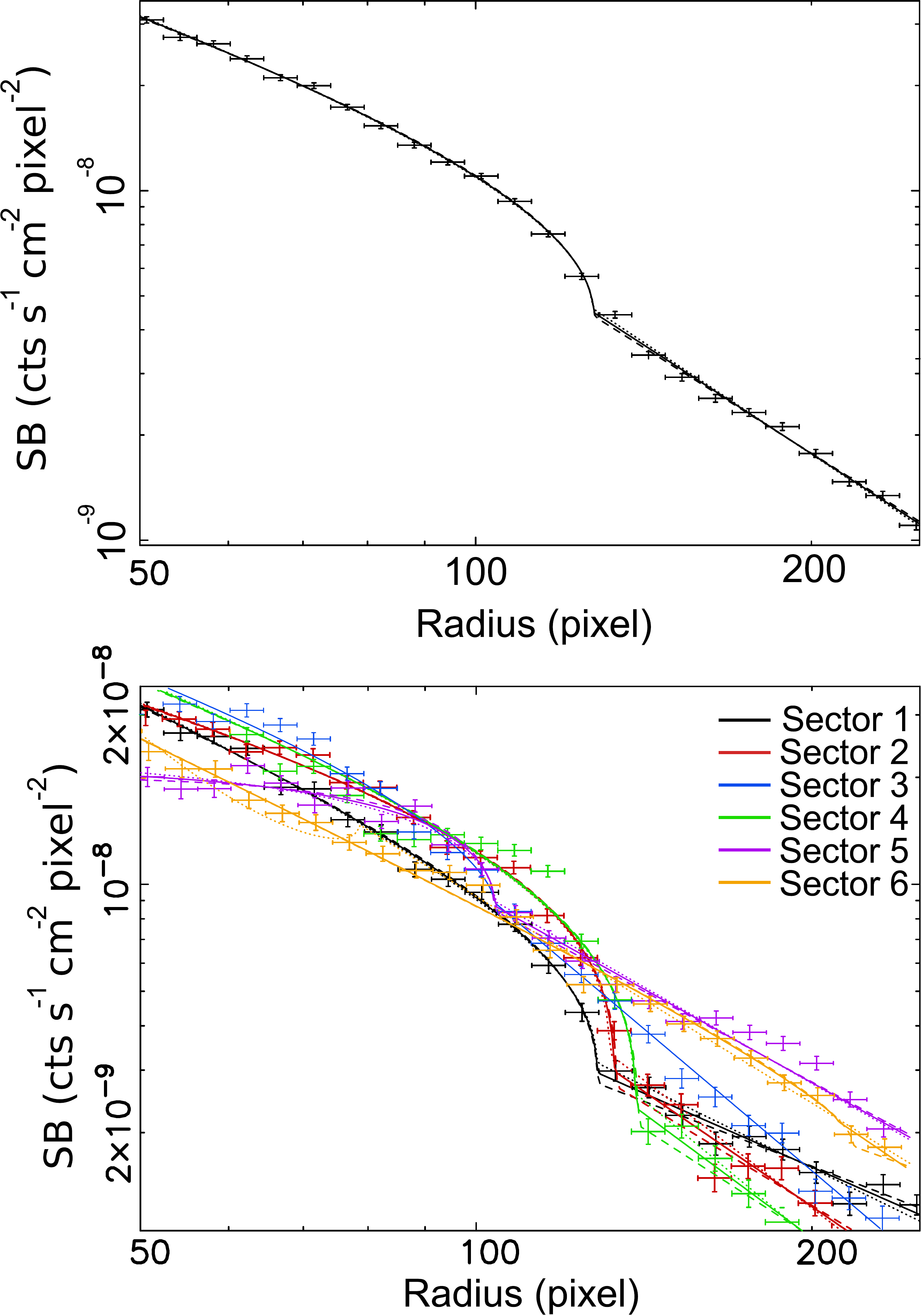}
\caption{\emph{Top:} The azimuthally averaged, 0.5--3~keV surface brightness profile across the inner surface brightness edge (between radii of 50 and 200 pixels or 6.9 and 27.5 kpc). The best-fitting broken power-law model is shown as the solid line. \emph{Bottom:} The 0.5--3~keV profiles and best-fitting broken power-law models across the inner surface brightness edge for the six sectors defined in Figure~\ref{F:shock_regions}.  A background of $7.9 \times 10^{-10}$ counts s$^{-1}$ cm$^{-2}$ pixel$^{-2}$ (determined from the double-$\beta$ model fit described in \S\ref{S:imaging_analysis}) was subtracted from both plots.}  \label{F:inner_shock_fits}
\end{figure}

To determine whether the surface brightness edge surrounding the cavities (see Figure~\ref{F:shock_regions}) is consistent with a shock discontinuity, we fit a broken power-law model to the azimuthally averaged surface brightness profile shown in Figure~\ref{F:inner_shock_fits}. In the fit, we chose the power-law indices inside and outside of the edge to match our observed surface brightness profile in those regions, and solved for the best-fitting radius and density jumpy for the discontinuity. As the power-law density profile is only an approximation to the true one, we fit over a limited range of radii near the location of the edge. Lastly, a background of $7.9 \times 10^{-10}$ counts s$^{-1}$ cm$^{-2}$ pixel$^{-2}$, determined by fitting a double-$\beta$ model plus a constant background to the surface brightness profile (see Table~\ref{T:beta_model}), was subtracted before fitting was performed.

The resulting best-fitting discontinuity parameters are given in Table~\ref{T:shocks} and the fit is shown in Figure~\ref{F:inner_shock_fits}. This discontinuity model fits acceptably (given the presence of significant non-azimuthally symmetric structure, such as the cavities, that distorts the surface brightness profile somewhat) and implies a density jump of 1.45 across the edge and a Mach number of 1.33.

However, if we fit to surface brightness profiles derived in subsectors (shown in Figure~\ref{F:shock_regions}), we do not find consistent fit parameters (see Table~\ref{T:shocks}).\footnote{We also find this behavior if elliptical regions, with ellipticity and position angle matched closely to those of the edge ($e=0.15$ and $\theta=28^{\circ}$ east from north), are used in the derivation of the surface brightness profile.} Although sectors 3 and 5 are clearly affected by the cavities and fits to the profiles in these sectors may not be meaningful, fits to the other sectors should be consistent with one another.

Instead, there is a large range in the best-fitting radii and density jumps among these sectors, with those in sectors 4 and 6 differing the most. Due to symmetry, one might expect that the fits in sectors 1 and 4, located on opposite sides of the core, would be similar; however, the best-fitting density jumps for these sectors differ at the 3-$\sigma$ level. Furthermore, in sector 6 there does not appear to be any significant edge in the surface brightness profile at the expected radius.

Although substructure in the core makes it difficult to obtain reliable fits, our analysis implies that a single discontinuity model cannot readily account for the surface brightness edge around the entire circumference. Furthermore, the inner temperature jump seen in Figure~\ref{F:profiles} corresponds closely to this edge and indicates that the temperature behind the edge is significantly lower than that beyond the edge. These results disfavor the shock interpretation. An alternative explanation for this edge is that it is due to the presence of rims associated with the cavities. We examine this possibility further in \S\ref{S:energetics}.

\begin{table}
\caption{Parameters of the Broken Power-law Models.}
\label{T:shocks}
\begin{tabular}{@{}lcccc}
\hline
 & \multicolumn{2}{c}{Break Radius} &  &  \\
Region$^a$ & (pixels) & (kpc) & Density Jump & $\chi^2_{\nu}$\\
\hline
   Inner edge, all sectors & 127.5 & 17.5 &  1.45 $\pm$ 0.04       & 2.39 \\
   Inner edge, sector 1    & 128.4 & 17.7 &  1.82 $\pm$ 0.17       & 1.32  \\
   Inner edge, sector 2    & 133.2 & 18.3 &  1.93 $\pm$ 0.18       & 2.12  \\
   Inner edge, sector 3    & 139.3 & 19.2 &  1.98 $\pm$ 0.18       & 9.28  \\
   Inner edge, sector 4    & 103.9 & 14.3 &  $1.21_{-0.76}^{+0.15}$& 2.39  \\
   Inner edge, sector 5    & 103.8 & 14.3 &  1.50 $\pm$ 0.10       & 3.89  \\
   Inner edge, sector 6    & 210.4 & 29.0 &  1.07 $\pm$ 0.4        & 2.01  \\
   Outer edge, sector 5    & 256.6 & 35.4 &  1.48 $\pm$ 0.17       & 1.11  \\
\hline
\end{tabular}
\medskip
$^a$See Figure~\ref{F:shock_regions} for region definitions.
\end{table}


\subsection{Energetics}\label{S:energetics}
Measurements of X-ray cavities have for the first time allowed a strong lower limit to be placed on the total jet power of AGNs. Figure~\ref{F:fullband_image}c shows an unsharp mask of the central region of HCG 62, showing the large cavities to the north and south of the core. These cavities are filled with radio emission \citep{gitt10}. There is also evidence of small inner cavities, oriented roughly north-south. The jet power of the outburst may be estimated from the enthalpies and buoyant rise times of the cavities.

To quantify the significance of the cavities, we can compare counts in the cavity regions shown in Figure~\ref{F:fullband_image} relative to those in undisturbed regions of the same size and at the same radius from the core. We find that the outer cavities represent deficits of $\approx 28$ per cent ($2062\pm 117$~counts) for the southern cavity and $\approx 14$ per cent ($1513\pm 142$~counts) for the northern cavity (this value includes the spur of bright emission in the northern cavity visible in Figure~\ref{F:fullband_image}c). For the inner cavities, visible in Figure~\ref{F:fullband_image}d, the deficits are $\approx 16$ per cent ($121\pm 37$~counts) for the southern cavity and $\approx 17$ per cent ($198\pm 39$~counts) for the northern cavity. Therefore, the significance of the inner cavities is much lower than that of the outer cavities, and, consequently, we examine the large outer cavities in much greater detail.

\begin{figure*}
\includegraphics[width=168mm]{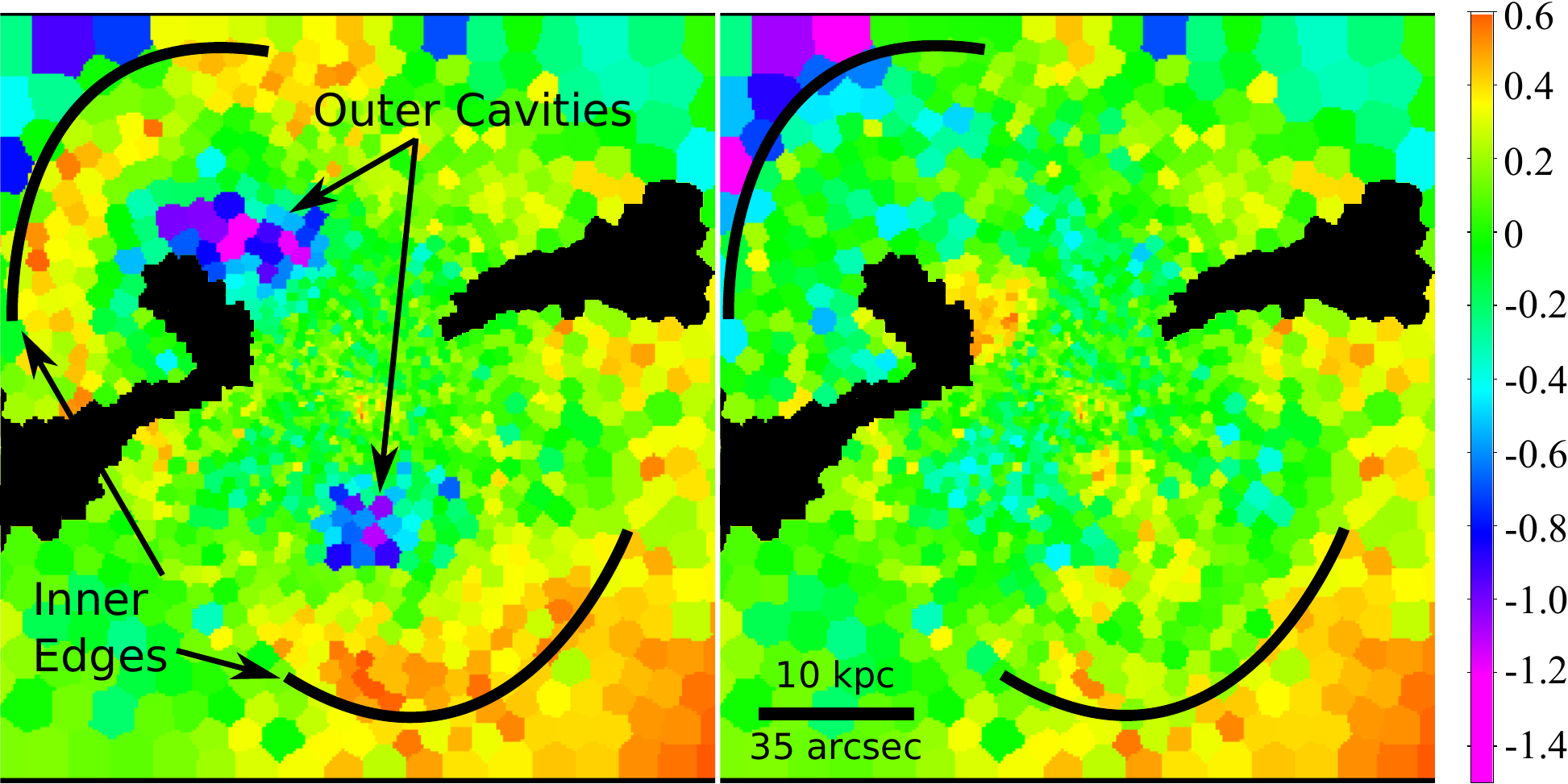}
\caption{Residual images (in the sense of [image - model]/image) of the core, made using the 0.5--3~keV image binned to a S/N of 25 per bin, for a model that includes only a 2-dimensional double-Beta component (\emph{left}) and for the best-fitting model that also includes empty spherical cavities surrounded by uniform spherical shells (\emph{right}). Masked regions are shown in black. The outer cavities and inner surface brightness edges are marked.}
\label{F:resid_image}
\end{figure*}

To make the best possible estimate of the energetics of the outer cavities, we used simulated images to recover the cavity sizes and locations. The large-scale cluster emission was modelled with the 2-D double $\beta$-model described in \S\ref{S:imaging_analysis}, the cavities as empty spheres, and the rims as uniform shells of emission that fully enclose the cavities. The 0.5--3 keV exposure-corrected image was binned using the WVT binning method of \citet{dieh06} to a S/N of 25 per bin. The same binning scheme was then applied to the model image at each iteration of the fitting process, which consisted of a least-squares fit performed with the TNMIN function of MPFIT \citep{mark09}. The spur of bright emission in the northern cavity was masked, as was the bright emission that runs between the cavities, perpendicular to the cavity axis (see Figure~\ref{F:fullband_image}c). A summary of the best-fitting cavity parameters is given in Table~\ref{T:cav_fit}, and the residual image is shown in Figure~\ref{F:resid_image}. For this fit, we obtained $\chi^2_{\nu} = 3.2$ (this includes contributions from non-azimuthally symmetric regions of the image, such as the arc of excess emission, that were not well fit by the 2-D double $\beta$-model).

It is clear from Figure~\ref{F:resid_image} that the cavity regions and inner surface brightness edges are well fit by the model. However, there remain residuals in the cavity regions that may indicate that they are distorted somewhat from spheres. In the northern cavity, some of the residuals may be due to the spur of bright emission noted above. The nature of this spur is unclear. One possibility is that it is cool material uplifted by the cavity in its wake, and, in this case, the bubble is probably nonspherical. However, deriving the true cavity geometry would be difficult due to projection effects and is beyond the scope of this work.

\begin{table}
\caption{Parameters of the Best-fitting Cavity Model. \label{T:cav_fit}}
\begin{tabular}{@{}lcccccc}
  & $r$ & $R$ & $\phi$$^a$ & $\theta$$^b$ & Rim width & \\
Cavity & (kpc) & (kpc) & (deg) & (deg) & (kpc) & $f_{\rm Rim}$$^c$ \\
\hline
   North & 5.5 & 9.3 &  90  & 135 & 5.5 & 0.89 \\
   South & 4.3 & 6.9 &  90  & 288 & 4.4 & 0.95 \\
\hline
\end{tabular}
$^a$Angle to the line of sight of a line connecting the cavity centre to the group core.\\
$^b$Angle in the plane of the sky, increasing counter clockwise north from west, for a line connecting the cavity centre to the group core.\\
$^c$Ratio of the total emissivity in the rims relative to the emissivity displaced by the cavities (i.e., a ratio of unity would imply equal emissivity in the rims and in the region displaced by the cavity).
\end{table}

Interestingly, the total emissivity in the rims of the best-fitting model is very close to that displaced by the cavities (within 10 per cent). This similarity could imply that the rims are composed of this displaced material. However, it could also be simple coincidence. A similar analysis for a sample of systems with cavities would be useful in understanding whether such a situation is common or not. We note that the inclusion of rims does not change the best-fitting cavity properties appreciably, but it does largely account for the inner surface brightness edges discussed in \S\ref{S:inner_sb_edge}. However, given the complex structure in the core, it is difficult to say conclusively whether the rim interpretation or the weak shock interpretation is the correct one (some combination of structure from both a rim and a shock is also possible).

The best-fitting values of the cavity sizes and locations give a total cavity enthalpy ($H = 4pV,$ where $p$ is the average pressure at the radius of the center of the cavity, calculated from the pressure profile shown in Figure~\ref{F:profiles}, $V$ is the volume, and assuming a relativistic plasma fills the cavities) for the outer cavities of $H_{\rm tot} \approx 3.6\times10^{57}$~erg (see Table~\ref{T:pv} for the derived cavity properties). This value is somewhat larger (by a factor of two) than that found by other studies \citep[e.g.,][]{birz04, mori06, gitt10}, although it is a factor of two smaller than the value derived by \citet{gitt10} using cavity sizes estimated from the radio emission. For the small inner cavities, we estimate (by fitting ellipses to them by eye) a total enthalpy of $\sim 1.1\times 10^{56}$~erg. Using the buoyant rise times as the ages ($t_{\rm buoy}$) of the cavities \citep[e.g.,][]{birz04}, we find a total mechanical power ($H_{\rm tot} / t_{\rm buoy}$) for all four cavities of $L_{\rm mech} = 8.6 \times 10^{42}$~erg~s$^{-1}$, a factor of $\sim 5$ above the cooling luminosity of $1.8 \times 10^{42}$~erg~s$^{-1}$ \citep{raff06}.

\begin{table*}
\caption{Cavity Energetics.}
\label{T:pv}
\begin{tabular}{@{}lccccccc}
 & $p$ & $V$ & $4pV$ & $t_{\rm buoy}$ & $4pV / t_{\rm buoy}$\\
Cavity & ($10^{-11}$ erg cm$^{-3}$) & ($10^{66}$ cm$^3$) & ($10^{56}$ erg) & ($10^7$ yr) & ($10^{42}$~erg~s$^{-1}$) \\
\hline
   Inner N & $9.7 \pm 1.1$ & 0.12 & 0.48 & 0.31 & 0.49\\
   Inner S & $9.7 \pm 1.1$ & 0.16 & 0.6 & 0.30 & 0.63\\
   Outer N & $2.9 \pm 0.2$ & 20.8 & 24.0  & 1.7  & 4.5\\
   Outer S & $3.3 \pm 0.2$ & 9.5  & 12.4  & 1.3  & 3.0\\
\hline
\end{tabular}
\end{table*}

\subsection{Cavity Contents}
Given the total radio luminosity of $3 \times 10^{38}$~erg~s$^{-1}$ \citep{gitt10}, the implied radiative efficiency (defined as $L_{\rm rad} / [L_{\rm rad} + L_{\rm mech}]$) is $\sim 4\times 10^{-5}$. Under the assumption of equipartition, the requirement of pressure support constrains the content of the lobes. For the observed radio luminosity, the relativistic electrons in the lobes can provide only a tiny fraction ($< 0.2$ per cent) of the particle pressure required \citep{mori06, birz08, gitt10}.

This result implies a large population of heavy particles (such as cosmic rays) or hot thermal gas in the lobes. For a given cavity geometry and background IGM distribution, one can use the surface-brightness residuals in the cavity regions to place limits on any thermal gas in the cavities \citep[e.g,.][]{schm02,mori06}. However, the true cavity geometry is difficult to constrain (as are deviations in the gas distribution from the beta model atmospheres assumed in most analyses), and arguments which rely critically on knowledge of the true cavity geometry and gas distribution may not provide useful constraints. For example, if the cavities are elongated along the line of sight (relative to spherical cavities), a thermal gas inside the cavities could be present but result in little to no residual emission when a spherical cavity model is used. To minimize these issues, we adopt a method similar to that used by \citet{sand07} that uses spectral constraints and depends much less on knowledge of the true cavity geometry.

To this end, we extracted spectra from circular regions of radius 6\arcsec\ in the deepest regions of the outer cavities, where any contribution from thermal gas inside the cavities relative to gas outside the cavities should be greatest. The spectra, responses, and backgrounds were made as described in \S\ref{S:spectral_analysis}. We fit the spectra with a model that includes two single-temperature components. The first component, with the temperature fixed to the average of that at the radius of the cavity, accounts for the projected emission. The second component accounts for any thermal emission inside the cavity, and its temperature is fixed to a value that was varied from 1.5~keV (the maximum temperature seen in the temperature profile of Figure~\ref{F:profiles}) to 5~keV. The metallicities were fixed to the the value obtained from deprojection at the cavity radius. Lastly, a WABS component was added to account for Galactic absorption, with the $N_{\rm H}$ column density fixed to $3.03 \times 10^{20}$~cm$^{-2}$, the value used throughout this paper. The full model is then WABS $\times$ (MEKAL$_1$ + MEKAL$_2$).

\begin{figure}
\includegraphics[width=84mm]{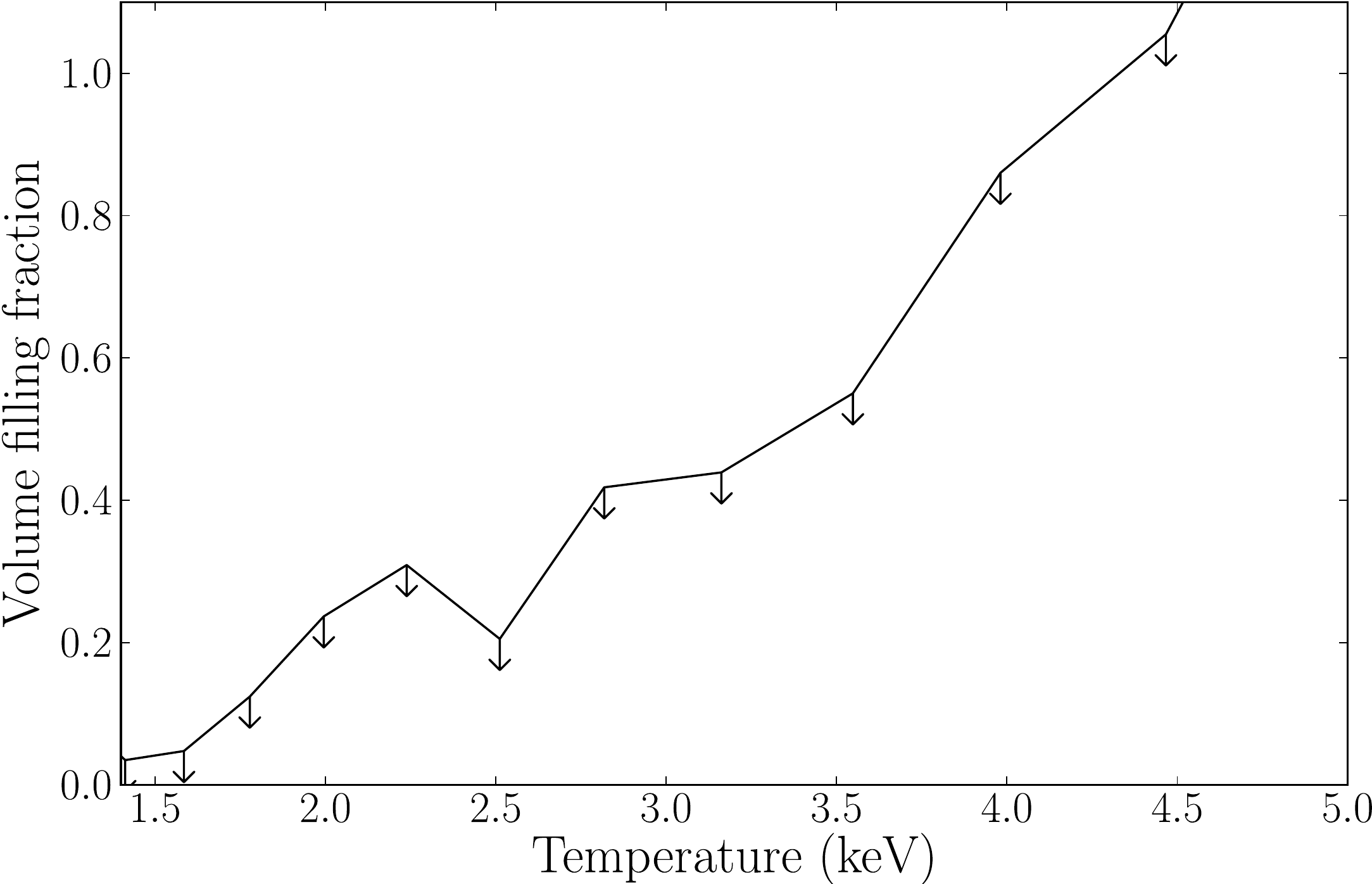}
\caption{The limit on the volume-filling fraction versus the temperature of any thermal gas in the southern cavity.}
\label{F:cavity_contents}
\end{figure}

Assuming that any emission from gas at temperatures above 1.5~keV comes from gas inside the cavity that is in pressure balance with the gas surrounding the cavity, we can place constraints on the volume-filling factor of this gas using upper limits on the normalization of the second MEKAL component. For the temperature and density of the surrounding gas, we use the deprojected values (from the deprojection analysis discussed in \S\ref{S:deprojection}) at the radius of the cavities. We assume that the cavities in the regions from which spectra were extracted are cylinders of length 11~kpc and 8.5~kpc for the northern and southern cavities, respectively (corresponding to the diameters recovered from our fits described in \S\ref{S:energetics}). We find that a volume-filling gas with a temperature less than 4.3 keV is excluded by our spectral fits in the southern cavity (see Figure~\ref{F:cavity_contents}). Fits to the northern cavity did not provide useful constraints (i.e., we could not exclude the presence of volume-filling gas of any temperature above 1.5~keV), probably due to the lower contrast any such gas would have at the larger projected radius of this cavity. We note that differences of 20 per cent in our adopted cavity diameter result in changes of $\approx 10$ per cent in the temperature limit (e.g., a diameter 20 per cent smaller results in a limiting temperature of $\approx 3.9$~keV).

Our limit of 4.3~keV on the temperature of volume-filling gas in the southern cavity is somewhat higher than the value of 3~keV determined by \citet{mori06} under the assumption of spherical cavities. However, it does not of course rule out the possibility that significant pressure in the cavities could be supplied by a gas with a temperature above this limit. Although deeper observations would allow this temperature limit to be increased, it is impossible to rule out gas at arbitrarily high temperatures (due to the scaling of emissivity with temperature for a gas in pressure balance with its surroundings).

\section{A Recent Merger}\label{S:merger}
In this section we examine evidence for a recent merger in the surface brightness distribution and temperature and metallicity structure of the IGM. In particular, we investigate whether the outer edge noted in \S\ref{S:imaging_analysis} is consistent with a weak shock (possibly generated by an AGN outburst) or rather with a cold front (generated by a merger), and we examine how the spiral-like excess of emission visible in Figure~\ref{F:fullband_image}b relates to this edge and to the merger scenario.

\subsection{The Outer Edge: Shock or Cold Front?}\label{S:outer_edge}
As noted by \citet{gitt10}, there is a clear surface brightness edge $\sim 2$ arcmin ($\sim 30$ kpc) to the south of the core. It was posited by \citet{gitt10} that this edge is associated with a weak shock, such as the one seen in MS0735+7621 \citep{mcna05}, that was possibly generated by the AGN outburst which created the cavities and radio lobes. In this section, we investigate whether this edge is consistent with a discontinuity formed by a weak shock and we examine the temperature structure of the gas near the edge.

We first fit the surface brightness profile across the edge with the broken power-law model discussed in \S\ref{S:inner_sb_edge}. We extracted the profile from the region of the outer edge only (sector 5, shown in Figure~\ref{F:shock_regions}), and, as before, we subtracted a constant background of $7.9 \times 10^{-10}$ counts s$^{-1}$ cm$^{-2}$ pixel$^{-2}$.

\begin{figure}
\includegraphics[width=84mm]{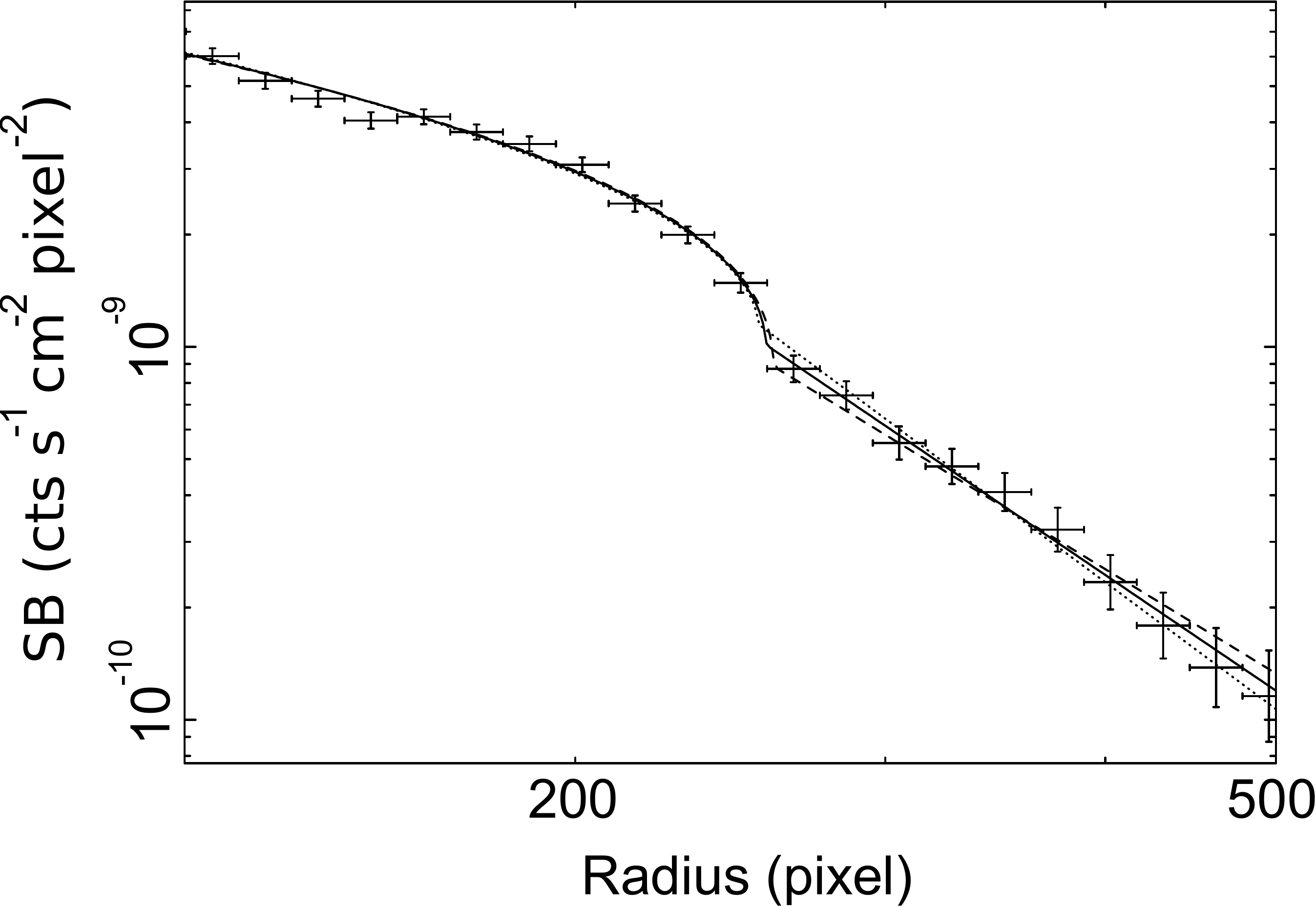}
\caption{Surface brightness profile (0.5--3~keV) and best-fitting shock model for the outer surface brightness edge (note that 100~pixels equals 13.8~kpc). A background of $7.9 \times 10^{-10}$ counts s$^{-1}$ cm$^{-2}$ pixel$^{-2}$ (determined from the double-$\beta$ model fit described in \S\ref{S:imaging_analysis}) was subtracted from the plot.}  \label{F:outer_shock_fits}
\end{figure}

We find a good fit to the observed profile (see Figure~\ref{F:outer_shock_fits}), with discontinuity parameters (see Table~\ref{T:shocks}) that match well with those found by \citet{gitt10} (although our best-fitting model implies a density jump $\approx 10$ per cent lower). Therefore, the edge in this region is consistent with the discontinuity expected to accompany a weak shock.

However, the temperature structure of the IGM in this region does not show the expected signature of a shock: namely, the temperature immediately behind the shock should be higher than that immediately in front. Instead, the temperature maps (see Figure~\ref{F:maps}) show that the temperature outside of the edge appears to be higher than that inside. Therefore, the temperature structure suggests that the discontinuity may be a cold front, and we examine this possibility further in the following section.

\subsection{The Spiral-like Excess and High Abundance Arc}\label{S:arc}
The arc of high metallicity gas noted in \S\ref{S:maps} was discovered in the original Chandra data by \citet{gu07} and was confirmed by a reanalysis of those data by \citet{gitt10}. In the new data, this arc is seen to correspond closely to a region of excess surface brightness (relative to the model of the large-scale emission described in \S\ref{S:imaging_analysis}; see Figure~\ref{F:maps}). The southern edge of this arc defines the outer surface brightness edge discussed in \S\ref{S:outer_edge}, and hence the arc and the edge appear to be related. Although a weak-shock model fits the edge well \citep{gitt10}, it is difficult to understand the temperature and metallicity structure of the arc in this context.

Alternatively, the arc and its edge could be indications of a recent merger, as suggested by \citet{gu07}. Indeed, these structures appear to match well with those seen in simulations of mergers in cooling flows \citep[e.g.,][]{asca06}. In this scenario, the merger induces sloshing in the cool core, uplifting material from the centre, and creating a low-temperature (and presumably high-metallicity) spiral-like feature and associated cold front (the outer edge). Additional support for the merger scenario in HCG 62 is provided by the kinematic study of \citet{spav06}, who found evidence that NGC 4778 has undergone a recent interaction, possibly with NGC 4761.

Therefore, although it is not possible to distinguish absolutely between the weak shock scenario and the merger scenario as the cause of the arc and surface brightness edge, the evidence obtained from the new \emph{Chandra} data appears to favor the merger scenario.

\section{Discussion}
\subsection{The Recent Merger}
We have presented evidence for a recent merger in the structure of the IGM. Our results suggest that the high-metallicity arc to the south of the core could be gas uplifted by sloshing induced by the merger. Alternatively, the high-metallicity arc material could have been stripped from the merging galaxy, as postulated by \citet{gu07}. Lastly, some of the metals in the arc could have been uplifted not by sloshing induced by the merger, but instead by the AGN, as seen in many clusters \citep{kirk11}.

However, the values of metallicity in the arc appear to exceed the central metallicity. It would therefore be difficult to understand how this gas could originate in the core. However, as we showed in \S\ref{S:spectral_analysis}, the true central metallicity could be higher than a single-temperature model indicates due to the presence of multi-temperature gas in the core. Indeed, the inclusion of a multi-temperature component (an MKCFLOW component) to fits in the core does result in central metallicities that are consistent with those in the arc. Therefore, in this scenario, the material in the arc could have been uplifted from the core, either by sloshing or by the AGN.

The simulations of \citet{asca06} show that sloshing is capable of uplifting cool material from the core and can result in structure similar to that seen in the surface brightness and temperature maps of HCG 62 (i.e., cold fronts and spiral-like arcs of lower-temperature material). However, it is not clear that AGN activity could easily replicate this structure, as such activity generally appears to distribute the metals along the cavity or radio jet axis in a fairly symmetrical way \citep[e.g.,][]{kirk11}. Furthermore, there is no evidence of enhanced metallicity in the direction of the northern lobe and cavity. Therefore, it appears unlikely that AGN activity alone could account for the observed metallicity distribution.

Whether the metals originate in the group core or were instead stripped from the merging galaxy is more difficult to constrain. However, we note that the  majority of material in the group with the highest metallicities would be expected to occur in the core where it would undergo the most enrichment. Additionally, there is no evidence of high-metallicity gas at the location of the probable merger candidate, NGC 4761, suggesting that the stripping must have been very efficient. It therefore seems probable that sloshing is the primary mechanism by which the high-metallicity arc was formed.

Such sloshing could have a significant heating effect on the cool core, as cool gas is mixed with hotter gas and turbulence increases \citep[e.g.,][]{shar09, zuho10}. However, in contrast to the cavities, any heating from such mechanisms is difficult to quantify. Furthermore, increased turbulence could conceivably affect entrainment of gas by the radio lobes, which might explain in part the high inferred values of the ratio ($k$) of the total particle energy to the energy in electrons for the outer cavities of HCG 62 \citep[on the order of hundreds;][]{birz08,gitt10}. In this scenario, one might expect that the inner cavities would have much lower values of $k$, as they would presumably have had less time to suffer the effects of entrainment. High resolution radio images of the inner cavities are needed to test this possibility.

\subsection{The AGN Outburst}
The cavity system in HCG 62 has been well studied, but there remain some poorly understood aspects. For instance, the deep radio image at 235 MHz presented in \citet{gitt10} clearly shows extensions of emission beyond the boundaries of the X-ray cavities. These extensions are spectrally steep, as they are not detected in the deep 610 MHz image of \citet{gitt10}. We have examined the regions where these extensions lie and find no evidence for significant X-ray structure that might be associated with them (such as older ``ghost'' cavities).  It is possible that such emission leaked from the cavities, as seen for example in M84 \citep{fino08}. However, given the small size of the extensions and their large distance from the core, the deficits produced by cavities would likely be undetectable in the current image.

If leakage of radio plasma out of the cavities did occur, leakage of thermal gas into the cavities might then be expected. Again, such a scenario is consistent with the high values of $k$ inferred for the cavities. If such gas has leaked into the cavities, it presumably has either been heated to temperatures above the limit of $\approx 4.3$~keV (for the southern cavity) or fills only a fraction of the cavity volume (e.g., $\approx 20$ per cent for a gas at 2~keV).

As for the energetics of the AGN outburst in HCG 62, we have shown that only $\approx 20$ per cent of the total cavity enthalpy is sufficient to balance cooling if it is injected on time scales similar to the buoyancy time scales inferred for the outer cavities. The remaining 80 per cent presumably goes (eventually) into net heating of the IGM. The net heating from the cavities is then at least $0.8H_{\rm tot} \approx 3\times 10^{57}$~erg. Given the total gas mass of $1.6\times 10^{12}$~$M_{\odot}$ within a radius of 300 kpc \citep{mori06}, the current outburst is supplying $0.8 H_{\rm tot} \mu m_{\rm p}/M_{\rm gas} \sim 6 \times 10^{-4}$~keV per particle of heat (where $m_{\rm p}$ is the proton mass and we have adopted a mean molecular weight, $\mu$, of 0.59). Assuming a typical inter-burst timescale of $\sim 1.5 \times 10^7$~yr and summing this heat over the life of the group ($\sim 7.7 \times 10^9$~yr, the lookback time to $z=1$), we find a total heating by the AGN of $\sim 0.3$~keV per particle.

This value is somewhat higher than the average value of $0.17$~keV per particle inferred by \citet{ma11} from the radio luminosities of a sample of 169 clusters. It is significantly lower than the heat required to break self-similarity \citep[$\approx 1$--3~keV per particle;][]{wu00}. However, the heating rate could well have been higher in the past, when the supermassive black hole at the core of the group was likely experiencing the bulk of its growth. Indeed, studies that estimate the average heating indirectly, such as from estimates of the black hole mass \citep[e.g.,][]{wang10} or from the gas profiles \citep[e.g.,][]{math11}, find much larger values of heating, implying higher heating rates at earlier epochs.

\section{Conclusions}
We present results of an analysis of new Chandra data of HCG 62. Our main findings are as follows:
\begin{enumerate}
\item We refine previous measurements of the AGN cavities and confirm the extremely low radiative efficiency of the outburst. This low efficiency constrains the pressure support from electrons in the cavities to be much less than needed to balance the external pressure of the IGM, implying that the cavities contain either thermal gas (possibly entrained by the lobes) or a significant population of cosmic rays.
\item Using spectral fits, we constrain the temperature of volume-filling thermal gas in the southern cavity to be $kT \gtrsim 4.3$~keV. However, gas at lower temperatures with a lower volume filling fraction is allowed by our fits.
\item For the first time, we detect the presence of an X-ray AGN at the core of HCG 62. We constrain the X-ray luminosity of the central AGN to be $L_{0.5-7{\rm ~keV}} = (1.5_{-1.0}^{+2.8})\times 10^{39}$~erg~s$^{-1}$ (3-$\sigma$ errors). The X-ray spectrum shows evidence of moderate absorption ($N_{\rm H} \approx 0.6 \times 10^{22}$~cm$^{-2}$).
\item We find evidence of a recent merger in the surface brightness, temperature, and metallicity structure of the hot gas. Such a merger may have contributed to the heating of the cool gas in the core, although the AGN cavities alone represent more than enough power to offset cooling.
\end{enumerate}
In the future, high resolution radio observations would be useful in verifying the inner cavities detected in the new data at low significance and in constraining their properties.

\section*{Acknowledgments}
We thank the referee for helpful comments that improved the paper considerably. DAR was supported in part by the NASA ADP grant NNX10AC99G and CXC grant GO9-0136a.

\bibliographystyle{mn2e}
\bibliography{/Users/rafferty/Documents/Bibliography/master_references}

\end{document}